\documentclass{article}

\usepackage{fullpage}
\usepackage{subfigure}
\usepackage{graphicx}
\usepackage{multirow}
\usepackage{url}

\def\clawpack{Clawpack~}
\def\manyclaw{ManyClaw~}

\title{ManyClaw:
Slicing and dicing Riemann solvers for next generation highly parallel architectures}

\author{
Andy R. Terrel \\
Texas Advanced Computing Center, University of Texas at Austin
\and
Kyle T. Mandli \\
Institute for Computational Engineering and Science, University of Texas at Austin
}

\date{}
\begin{document}

\maketitle

\let\thefootnote\relax\footnotetext{We would like to thank W. Nathan Bell of NVidia Corp. for an initial
  TBB implementation, Gorune Ohannessian of the American University of Beirut for an
  initial advection solver implementation, and King Abdullah
  University of Science and Technology for hosting the [HPC]$^3$ Conference
  where this work was initiated.}

\begin{abstract}
     Next generation computer
    architectures will include order of magnitude more intra-node parallelism; however, many
    application programmers have a difficult time keeping their codes current
    with the state-of-the-art machines. In this context, we analyze Hyperbolic
    PDE solvers, which are used in the solution of many important
    applications in science and engineering. We present \manyclaw, a project
    intended to explore the exploitation of intra-node parallelism in hyperbolic
    PDE solvers via the \clawpack software package for solving hyperbolic PDEs.
    Our goal is to separate the low level parallelism and the physical
    equations thus providing users the capability to
    leverage intra-node parallelism without explicitly writing code to
    take advantage of newer architectures.
\end{abstract}

\section{Introduction}
As the scientific computing community approaches the hurdles necessary to reach
exascale computation, many are looking at how current PDE solver packages can
be easily transformed for the coming changes. Current practices require heavy
involvement from application scientists and engineers in the algorithm and
design process. Lowering such barriers will facilitate the adoption of
next-generation computer architectures for a wider number of applications. Many
of the current solvers require the use of methods such as adaptive mesh
refinement, that make the task of identifying data exchanges for not only MPI
tasks but now in the context of intra-node parallelism which requires the
modeling of shared data and execution on coprocessors.

In this paper we propose and analyze a number of different strategies for
intra-node parallelism in the context of hyperbolic PDE solvers via
ManyClaw\cite{ManyClaw}. ManyClaw is a many-core implementation of \clawpack
\cite{Clawpack:tp}, a popular and representative finite volume package that has
a large number of applications ranging from tsunami propagation across the
Pacific to elasticity problems in complex material including the human body. We
hope to provide a number of different approaches to leveraging intra-node
parallelism analyzing each for their scalability and suitability for easy
adaptation by application developers.

\subsection{Hyperbolic PDE Solvers and Finite Volume Methods}

Hyperbolic PDEs form the basis of many important physical phenomena that are of
interest to a wide variety of scientists and engineers. The most basic
hyperbolic PDEs include the linear wave equation and itself capture many
important processes. These linear hyperbolic PDEs are often approximations to
fully nonlinear systems of PDEs which can exhibit complex behavior such as the
development of discontinuities from smooth initial conditions in finite time
(commonly called shocks). Solvers that attempt to solve these nonlinear systems
are often costly and require the use of sophisticated nonlinear computational
kernels such as Riemann solvers and limiters. \clawpack is one such solver
employing a finite volume wave-propagation algorithm that includes these
computational kernels and provides a proven approach to solving hyperbolic PDEs
in general.

Finite volume methods discretize a domain by partitioning it into cells and
then evolving the average value of each quantity of interest. At the edges of
these cells a flux must be calculated which often involves the computation of
the solution to the Riemann problem which itself is composed of the original
PDE on an infinite domain with a discontinuity in the initial conditions at
$x=0$. The simplification of the general problem into a series of simpler
problems is one of the core advantages of finite volume types of methods and
allow for the physics of the problem to be largely decoupled from the algorithm
and computational method being used. For this reason, \manyclaw currently
concentrates on implementing the solution of a variety of Riemann solvers using
multiple different approaches to intra-node parallelism with the requirement
that the underlying Riemann solver does not have to be aware of the parallelism
and is left unchanged from a serial execution of the Riemann solvers.

\subsection{Computational Decomposition}

The kernels used to evolve a solution forward in time can be mapped by
arithmetic intensity to various levels of a heterogeneous compute architecture.
In the case of \clawpack, the most intense kernels including the Riemann solver
and limiter routines can be mapped to coprocessor technologies, mid-level
kernels such as local time step management and flux accumulation to the
multicore node, and low intensity kernels such as grid management to the
distributed set of nodes. For this paper we concentrate on only the high
intensity Riemann solver routines as this approach has the potential to
significantly accelerate many application developer codes with minimal
intervention on their part.

Some threads will handle larger portions of the data and
dynamically schedule pieces to finer-grained threads. Such a dynamic
execution model, matching exascale machine designs, allows for local precision
with minimal effects to global time stepping.

\begin{table}
  \begin{tabular}{|cccc|}
    \hline
    Kernel & Algorithmic Intensity & Data in & Data out\\
    \hline
    Grid Management & low & grid and refinement details & refined grids \\
    Dynamic Time stepping & low to mid & time step of adjacent grid & updated
    values at grid boundaries\\
    Riemann Problem & high & grid cell centers & values at grid boundaries \\
    Limiters & high & values at grid boundaries & grid cell centers \\
    \hline
\end{tabular}
\label{table:explicit_kernels}
\caption{Summary of kernel for explicit methods for hyperbolic PDEs}
\end{table}

Grid management must marshal data between processes running different portions
of the grid. Additionally, it must manage the refinement between grids and
occasionally adjust the load balance between processes. Finally, it serves as
the dynamic dispatcher for the time stepping kernels. As the intra-node
parallelism is increased, this operation will be managed even between devices
and shared memory pools. For the most part, this problem has been
solved~\cite{PyClaw:te,Ketcheson:2011tj} and scales to leadership-class
machines. Thus we leave its analysis for other work.

To mitigate the CFL condition limits, each refined grid can manage its own time
step. This requires an update procedure on refinement boundaries which varies
by order of spacial or temporal interpolation. These operations will have to
grab data from the values produced by fine grained threads, such as those on
a coprocessor, and work in its own node level thread.

The Riemann problem implements the updates of the cell fluxes from field values
based upon the specific modeling equations. This is the most arithmetic intense
portion of the code and is dependent on the equations being solved. The minimal
overlap between cells lends it to be highly amenable to coprocessors.

The final kernel to discuss is the limiter which updates the field values based
upon the fluxes computed by the Riemann problem.  These simple operations
require more cell data in the computation but is still local to the Riemann
problem. Limiters also must match up between different sizes of grids requiring
extra overhead associated with the mesh refinement.

\section{Threading the Riemann solver}

The rest of this preliminary report will focus on the scalability of threading
different Riemann problems. Two recent debates in the \clawpack community
guided this study. The first issue is the type of code that must be written for
speed. While the numerical scientist thinks about the Riemann problem most
naturally in a pointwise manner, i.e. considering only a single interface at a
time, the \clawpack code used vectorized Riemann problems presumably for speed.
The second issue regarded the CUDA implementation of \clawpack that has not
been adopted by the community.  While CUDACLAW~\cite{CUDAClaw} showed speedups of up to 40X
over the \clawpack implementation, it did so at considerable code complexity.
To keep the decomposition easy to reuse by the community, we tried to keep the
work in standard work arrays that in practice will cause bandwidth bottlenecks.
The resolution of this second issue has been left as further work.

To address these issues, we implemented the four standard Riemann problems with
varying arithmetic intensity via vectorized Fortran, OpenMP, and TBB.  The code
was factored to allow any threading model to call a Riemann problem, provided
for by the \clawpack community, over a structured grid.  Our hope is to add a
more threading models, such as Intel SPMD Program Compiler and Thrust CUDA, to
further improve the study.

The main function evaluated is the Riemann problem, which takes the states and
auxiliary variables of two or more adjacent cells and computes the flux between
the cells.  The simplest loop is to compute this at each cell interface for
each direction one at a time.  We also implemented versions that compute each
cell direction at the same time and then one that tiles the data for better
cache locality.  The simplest and most efficient turned out to be calculating
each direction at the same time.  This was the basis for the threading models.

\subsection{Fortran Code Comparison}

The Fortran results are meant as a representative sample of how the original
\clawpack code performs. A number of different comparisons were produced based
on the Fortran code that is currently a part of \clawpack. The \clawpack
algorithms are currently implemented as a vector evaluation of a strip of the
discretized domain and sweeps are done in each direction. Three different
methods of access to the entire grid were implemented in an effort to
understand the access overhead that is imposed by the artificial access
patterns required for the tests. The first is the nearest implementation to
what is currently done in \clawpack, a direct access slice through the strided
arrays. The second uses Fortran pointers in order to window into the
appropriate data. The last is an artificial test that removes all overhead due
to accessing strips of the array. The last implementation reported is a
point-wise evaluation of the Riemann solver similar to what has been
implemented in the other tests.

\subsection{OpenMP parallel for}

The OpenMP code used C++ and the {\tt parallel for} construct over the cell
centers. For our tests, this code simply adds a pragma statement to the outermost
cell iterator.  In practice, when there are many threads managing the
time stepping and other grids we expect memory affinity and thread management
to degrade performance.  Our numbers in this since give us a best-effort
performance when there are no other factors of a highly parallel code running.

\subsection{Thread Building Blocks}

The Thread Building Blocks (TBB) implementation used C++, as required by the tool,
and the {\tt parallel\_for} execution over a {\tt blocked\_range2d} iterator.
TBB uses various iterators to smartly schedule the running kernel.  In our
results we see that this strategy which often requires some warm-up time to
schedule effectively works well for larger thread pools and problem sizes.

\section{Numerical Tests and Results}

Our numerical tests focused on four common Riemann problems and scalability on
both a 12 core Westmere and a 30 core Knight's Ferry supporting 120 threads.
Both these architectures were used at the Texas Advanced Computing Center.  The
Westmere is a commodity architecture similar to what many in the \clawpack
community use.  While \clawpack does not take advantage of intra-node
parallelism, it is understood to scale on coming highly parallel architectures
the Knight's Ferry is a good test ground.

The four kernels implemented are advection, constant coefficient acoustics,
variable coefficient acoustics and Euler. We have a rough calculation of the
arithmetic intensity of each kernel in Table~\ref{table:rp}, this number is
based on hand counting the stores and flops in source and is a rough order of
magnitude number.

\begin{table}
        \centering
\begin{tabular}{|lc|}
\hline
Riemann Problem & Arithmetic Intensity (flops/byte) \\ \hline
advection &  1 / 3 \\
constant coefficient acoustics &  4 / 5\\
variable coefficient acoustics & 1 \\
Euler & 1 \\ \hline
\end{tabular}
\caption{Riemann problem descriptions}
\label{table:rp}
\end{table}

\begin{table}
\centering
\begin{tabular}{|c|cccc|ccc|}
\hline
& \multicolumn{4}{|c|}{Fortran} & \multicolumn{3}{|c|}{C++} \\ \hline
& Direct Access & Pointers & Artificial & Point-wise & Row-wise & Cell-wise & Tiled \\ \hline
Advection & 162.2 & 245.7 & 35.8 & 56.6 & 313.9 & 74.9 & 80.9 \\
Euler & 1320.5 & 1478.4 & 1075 & 781.2 & 1338.0 & 934.1  & 934.5 \\ \hline
\end{tabular}
\caption{Runtimes for Fortran and C++ on 2048x2048 grid}
\label{table:fortvc++}
\end{table}

In general, the C++ kernels outperformed the \clawpack Fortran kernels but not
the rewritten point-wise Fortran, see Table~\ref{table:fortvc++}.  These
numbers were surprising to the authors, but the Fortran array indexing seems to
be causing the difference. The artificial test shows that calling the Riemann
problem was not the major overhead.

The threading tests are not quite conclusive and show several interesting
characteristics.  We show the timing and scaling of the runs with the Westmere
to give the reader a feel of how long these steps take.  Each millisecond
counts as it is quite often that the time stepper will require millions of
timesteps.  We show the advection (Figure 1) and Euler (Figure 2) problems as
they do the least and most amount of work, respectively.  While the timing data
is pretty smooth for both, the scalability jumps around a bit.  Perhaps the
most notable trend is the different scalability trends the Euler problem takes
based on the grid size.  Additionally TBB does worse on the advection kernel
presumably due to the lack of warm up time for scheduling.

Figure 3 shows the scaling of the code running on the Knight's Ferry.  It
continues to get about 50\% scaling for Euler and 40\% for advection, despite
the large number of cores. This bodes well for the programmer who can scale the
code on a current core such as the Westmere.

\begin{figure}[ht]
\centering
\subfigure{
\includegraphics[scale=.39]{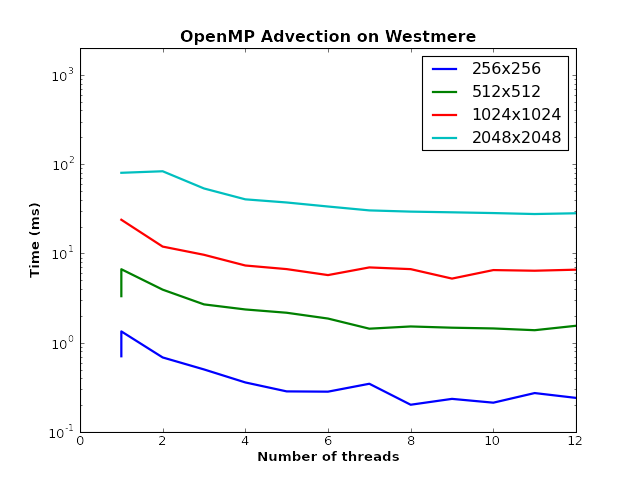}
\label{fig:subfig1}
}
\subfigure{
\includegraphics[scale=.39]{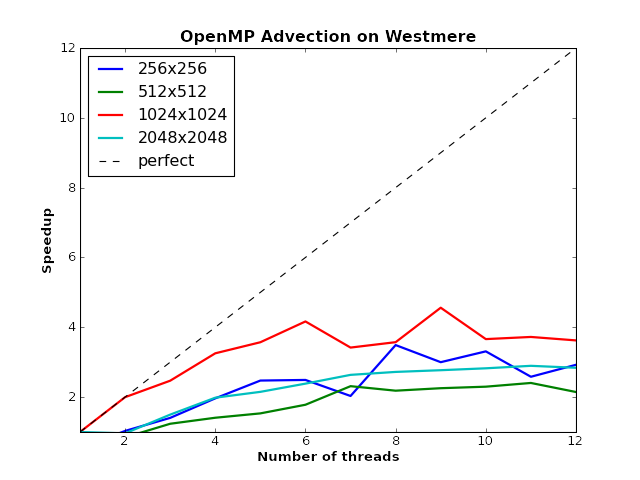}
\label{fig:subfig2}
}
\subfigure{
\includegraphics[scale=.39]{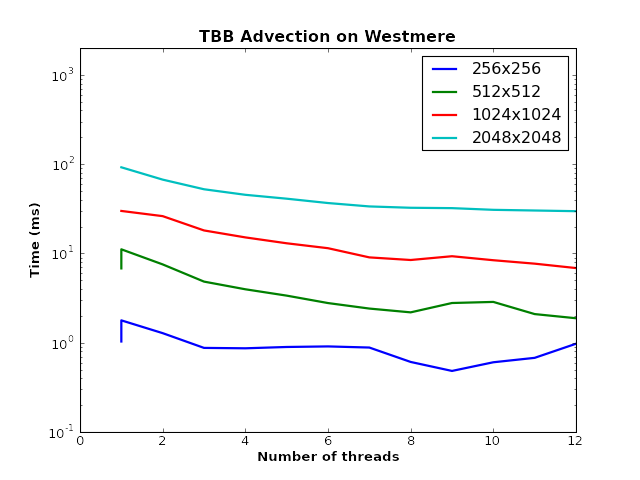}
\label{fig:subfig3}
}
\subfigure{
\includegraphics[scale=.39]{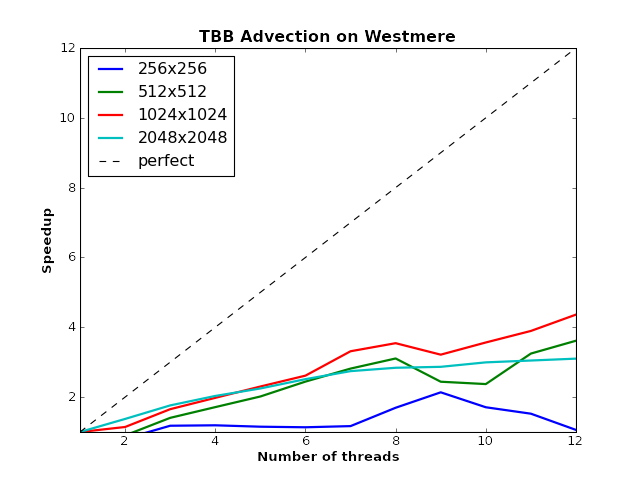}
\label{fig:subfig4}
}
\label{fig:subfigureExample}
\caption{Timing and Scaling of different grids on advection problems using Westmere.}
\end{figure}

\begin{figure}[ht]
\centering
\subfigure{
\includegraphics[scale=.39]{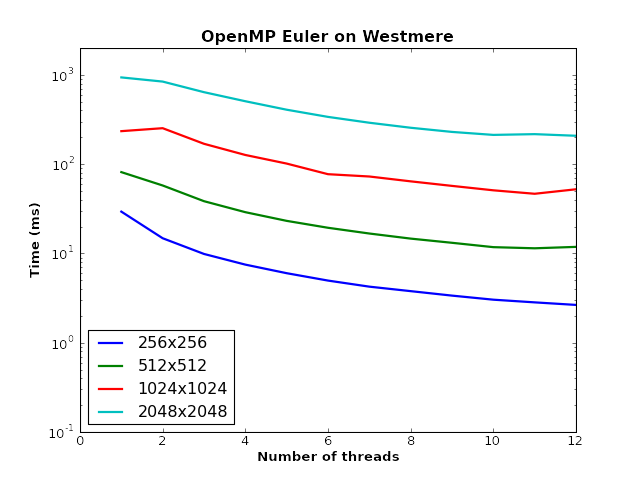}
\label{fig:subfig1}
}
\subfigure{
\includegraphics[scale=.39]{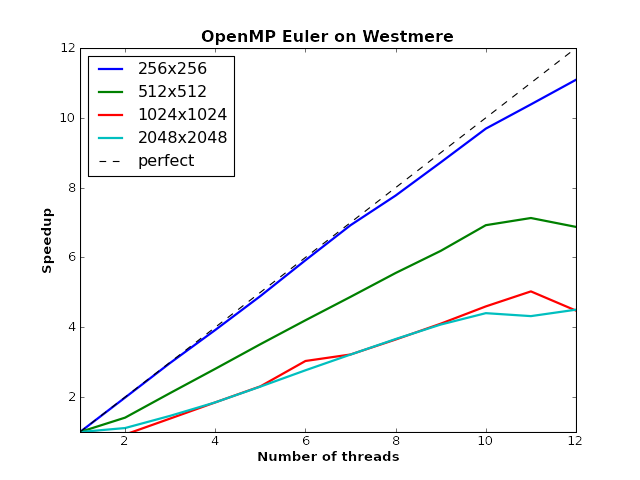}
\label{fig:subfig2}
}
\subfigure{
\includegraphics[scale=.39]{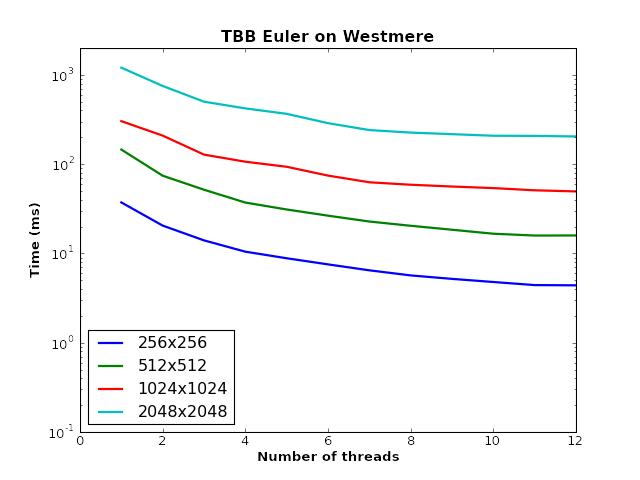}
\label{fig:subfig3}
}
\subfigure{
\includegraphics[scale=.39]{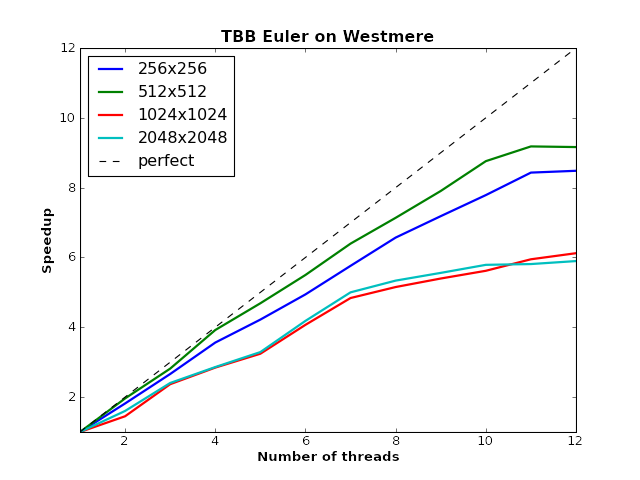}
\label{fig:subfig4}
}
\label{fig:subfigureExample}
\caption{Timing and Scaling of different grids on Euler problems using Westmere.}
\end{figure}

\begin{figure}[ht]
\centering
\subfigure{
\includegraphics[scale=.39]{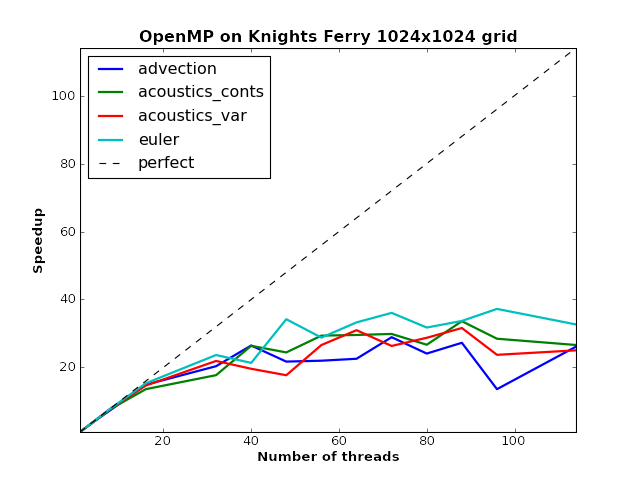}
\label{fig:subfig3}
}
\subfigure{
\includegraphics[scale=.39]{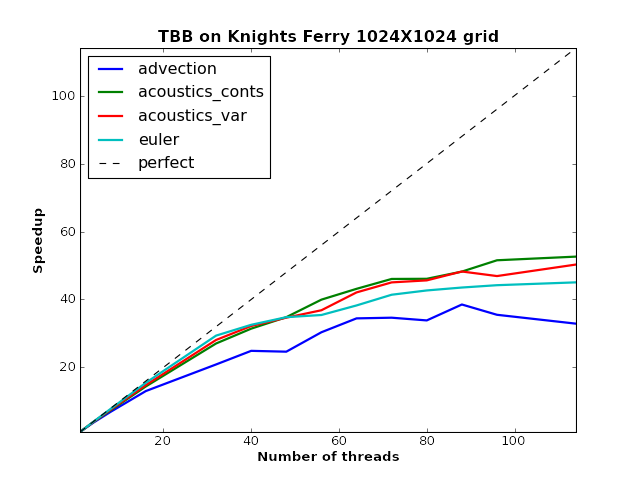}
\label{fig:subfig4}
}
\label{fig:subfigureExample}
\caption{Scaling of various problems using Knight's Ferry on a 1024x1024 grid.}
\end{figure}

\section{Conclusion}

In the paper we demonstrated an easy-to-use system that allows \clawpack users
to utilize intra-node parallelism with reasonable scaling results.  The work is
far from done as there still needs several parts of the algorithm to complete.
Notably, limiters, dynamic time stepping, and automatic mesh refinement.
Additionally, we expect to explore several other libraries for parallelism,
especially Thrust, Intel SMPD Program Compiler, OpenCL, and CUDA.

\bibliographystyle{plain}
\bibliography{bibliography.bib}

\end{document}